# Feature-preserving interpolation and filtering of environmental time series

Gregoire Mariethoz[1,2], Niklas Linde[1], Damien Jougnot[1,3], Hassan Rezaee[4]

[1]*Faculty of Geosciences and Environment, University of Lausanne, Switzerland*

[2]*School of Civil and Environmental Engineering, University of New South Wales, Sydney, Australia*

[3]*Sorbonne Universités, UPMC Univ Paris 06, CNRS, EPHE, UMR 7619 METIS, Paris, France*

[4]*Department of Civil, Geological and Mining Engineering, École Polytechnique de Montréal, Canada*

## Abstract

We propose a method for filling gaps and removing interferences in time series for applications involving continuous monitoring of environmental variables. The approach is non-parametric and based on an iterative pattern-matching between the affected and the valid parts of the time series. It considers several variables jointly in the pattern matching process and allows preserving linear or non-linear dependences between variables. The uncertainty in the reconstructed time series is quantified through multiple realizations. The method is tested on self-potential data that are affected by strong interferences as well as data gaps, and the results show that our approach allows reproducing the spectral features of the original signal. Even in the presence of intense signal perturbations, it significantly improves the signal and corrects bias introduced by asymmetrical interferences. Potential applications are wide-ranging, including geophysics, meteorology and hydrology.

## Highlights

- A time series processing method that repairs both gaps and interferences
- Preserves dependencies between multiple variables
- Quantifies the uncertainty of the reconstructed signal

# 1. Introduction

Time series are used universally in earth sciences as they constitute the most common output of measuring devices in disciplines such as hydrology, geophysics or meteorology. Almost as ubiquitous as the use of time series is the occurrence of periods with signal interference and measurement gaps. Such failures often constitute major impediments to quantitative analysis of environmental monitoring data. Examples include, among other fields, in-situ measurements of hydrological processes using geophysical methods (Smith-Boughner and Constable, 2012) or rainfall measurements (Kim and Ahn, 2009). This paper proposes a filtering method that addresses such contamination of time series data while preserving the underlying signal properties.

While well-established techniques exist to filter interferences and measurement errors, more challenging situations are intermittent interferences (e.g., measurement perturbations, earthquakes, magnetic storms) or data gaps (e.g., temporary measurement failure). Advanced time-series analysis, for instance time-frequency analysis with wavelets, often require continuous data for the entire measurement period, which implies that data gaps need to be filled by interpolation. In other applications, it is common to simply discard data from time-periods with significant interference. This excludes the possibility of using information in the underlying signal that could possibly be exploited. The focus of the present contribution is on signal interference and data gaps. To be effective, the ideal filtering approach should (a) maintain the same type of features as in measurement periods that are unaffected by data gaps or interferences; (b) recover information about the underlying signal that is hidden in the periods affected by interference; (c) provide uncertainty bounds relative to the filtered outputs and (d) preserve the coherence between several concurrent data sources. The last point can be illustrated by the example of two correlated time series, one of them showing low frequency variations and the other one high frequency fluctuations. If both time series are affected by a data gap, the reconstructed values need to preserve their joint statistical relationship, while maintaining the inherent variability specific to each time series.

The approaches commonly used for gap-filling and denoising include linear or spline interpolation and inverse weighted distances (Teegavarapu and Chandramouli, 2005). Spectral decomposition methods have been applied to time series with gaps (Schoellhamer, 2001), including multivariate data (Kondrashov and Ghil, 2006) and space-time domains (Wang et al., 2012). Parametric approaches such as autocorrelation models (Broersen, 2006) and kriging, which uses a variogram as statistical model of variability (Ruiz-Alzola et al., 2005), provide uncertainty estimates, but the resulting estimation is smooth and may not preserve the expected variability of the unsampled signal. Žukovič and Hristopulos (2013) introduced a directional gradient-curvature method to fill gaps in spatial remote sensing data. Recently, copula-based methods have been shown to outperform kriging for gap-filling problems (Bárdossy and Pegram, 2014). In another vein, Paparella (2005) formulates the gap-filling problem as an optimization that starts with a stitching of pieces from the observed signal.

In recent years, a family of non-parametric methods has emerged in geostatistics that are based on the recognition that parametric models may be poorly adapted to represent complex phenomena. Among these, multiple-point geostatistics use a statistical model of variability that consists of an example of the phenomenon studied, known as training data, from which high-order statistics or patterns are borrowed. The use of training data for filtering complex time series has not yet been investigated. We propose herein an approach based on training data consisting of the interference-free parts of the signal. Our approach performs well in the typical case where a large portion of the time series is unaffected by data gaps and interferences. In addition to populating the problematic periods with realistic data values, our approach is stochastic and allows for uncertainty characterization of the filtered results through multiple realizations.

## 2. Methodology

Our approach consists in iteratively replacing missing or interference-affected values by patterns sampled from the training data (i.e., recorded time-series that are unaffected by interferences or gaps). The sampling of the training data is achieved through the Direct Sampling algorithm (DS) that can perform multivariate simulation of continuous and/or categorical variables (Mariethoz et al., 2010). This algorithm was modified to perform iteratively, such that every iteration removes some of the interferences until convergence. The procedure adopted is illustrated in Fig. 1 and is described below. It is also available in a pseudocode form in Supplementary Table 1.

The input to the filtering algorithm is a noisy, possibly multivariate geophysical time series $\mathbf{S}_v(t)$, with $t$ denoting the time stamp and $v=1…V$ the variable considered. Firstly, all problematic time stamps affected by either interferences or data gaps are identified (denoted $t_{SIM}$), while the remaining time stamps are used as training data (denoted $t_{TD}$). The way of identifying noisy periods is case-specific (see application below). Each filtering iteration updates the values of the problematic time stamps and leaves unchanged the data that are unaffected by interference or gaps. These problematic time stamps can correspond to either gaps or interferences, however in practice all gaps are filled after the first iteration. All problematic time stamps, for all variables, are visited according to a random order that is initialized at each iteration. For each time stamp $t_{SIM}$ to be simulated, the data pattern **N** formed by the *n* closest informed neighbors of $t_{SIM}$ is identified (i.e. not including neighbors corresponding to interferences or gaps). The training data set is then searched in a random fashion for data patterns $\mathbf{N}_{TD}$ that are similar to **N**. As soon as one pattern $\mathbf{N}_{TD}$ is found that is sufficiently similar to **N**, its central value $S_v(t_{TD})$ is used to update the value currently considered. The random order in which time stamps are simulated and the random scanning of the training data ensure stochasticity in the process, hence allowing to generate multiple realizations.

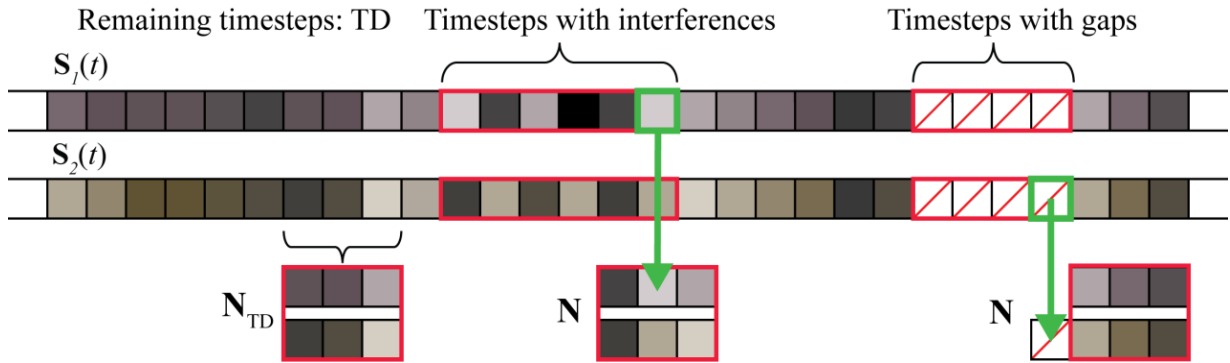

**Figure 1**. Illustration of the filtering algorithm for a bivariate case with one area affected by interferences and another area affected by gaps. For each case, the time stamp currently simulated is highlighted in green. The neighborhood size is *n*=3 for each variable, resulting in the 3 closest neighbors being chosen to simulate each time stamp (including the simulated time stamp itself in the case of interferences). The extracted neighborhood N is then compared to neighborhoods $\mathbf{N}_{TD}$ in the training dataset.

In interference-affected periods, **N** will initially present features that do not correspond to the patterns found in the training data set. With additional iterations, the patterns present in those interference-contaminated parts of the data become increasingly similar to the training data. Because the initial state contains the interference-affected data, some of the embedded signal characteristics are preserved. This is in contrast with the commonly used approach that consists in removing the interference-affected data and resimulating it without any possibility to recover useful signals properties within those gaps.

In the case of data gaps, the pattern matching procedure is used to fill the gap at the first iteration (expanding the neighborhood until $n$ informed values are found for each variable). In subsequent iterations this value is updated in the same way as for interference-affected values. In the sampling procedure, a training data value is accepted as the first occurrence where the distance $d(\mathbf{N}, \mathbf{N}_{TD})$, is smaller than a user-defined threshold value $\tau$. This "first matching wins" approach yields according to Shannon (1948) a statistically correct sampling of the conditional probability distribution corresponding to $\text{Prob}\{S_v(i) | \mathbf{N}\}$. The parameter $n$ defines the size of the data patterns considered, and also the order of the signal statistics to be reproduced. Values between 5 and 20 neighbors are recommended.

The data patterns $\mathbf{N}$ and $\mathbf{N}_{TD}$ can span over multiple variables, in which case $\mathbf{N}$ is a concatenation of the neighbors for each variable considered. Hence the data pattern comparison extends over many variables that are filtered simultaneously. This allows preserving linear or non-linear dependences between the variables present in the training data set.

Central to the method is the notion of similarity between patterns, defined by a distance function $d(.)$ that compares patterns from the problematic data with the training data. Since geophysical phenomena can present significant non-stationary behavior, it is expected that the underlying signal displays both small-scale and large-scale fluctuations. To account for this, we consider a distance measure whereby two data patterns are similar if they show the same fluctuations around their mean value, regardless of this mean value. We therefore use the following variation-based distance:

$$d(\mathbf{N}, \mathbf{N}_{TD}) = \frac{1}{n} \frac{|(\mathbf{N} - \overline{\mathbf{N}}) - (\mathbf{N}_{TD} - \overline{\mathbf{N}_{TD}})|}{max[\mathbf{S}_v(t_{TD})] - min[\mathbf{S}_v(t_{TD})]}, \qquad (1)$$

where $\overline{\mathbf{N}}$ denotes the average value in the pattern $\mathbf{N}$. When several variables are filtered simultaneously, we use a linear combination of the individual distances for each variable. The normalization term makes sure that the distances are dimensionless and can be compared across different variables.

## 3. Application to Geophysical Signal Processing

### 3.1. Presentation of the test data set

The new filtering method was applied to self-potential time series acquired at the Volund agricultural test site of the Danish hydrological observatory HOBE (www.hobe.dk). This example was chosen because it allows testing our method in the most challenging conditions as 1) it displays interference effects that are several order of magnitude larger than the underlying signal, 2) several interdependent time series are processed simultaneously and 3) the gaps and interferences are simultaneous for each time series, which means that there is no information at all within the affected time periods.

The self-potential method is a passive geophysical method in which spatial and temporal variations of the natural electrical field are measured. These variations are sensitive to different hydrological processes, such as water flux, ionic diffusion or redox phenomena (e.g., Revil and Jardani, 2013). Self-potential data consist in time series of electrical potential differences between a measurement and a reference electrode: $SP(t) = \varphi(t) - \varphi^{ref}(t)$, where $\varphi(t)$ indicates electrical potential [Volts] as a function of time $t$. In the present data set (Jougnot et al., 2015), the measurement electrodes are buried 3.1 (electrode 1), 2.5 (electrode 2), and 1.9 (electrode 3) m below the ground surface, while the reference electrode is located at 1.45 m depth. Data were recorded every 5 minutes from July 19, 2011, to October 18, 2012. Due to technical reasons, the acquisitions were frequently interrupted during time periods of minutes to several days. Furthermore, repeated cross-borehole electrical resistivity tomography measurements at the site caused strong interference. Indeed, the potential distribution caused by the artificially injected currents is orders of magnitudes more important than the natural ones of interest. These periods of interference are known or they can be identified in the time series by high-magnitude and high-frequency signal characteristics. The self-potential data, with its noise characteristics, are displayed in Fig. 2. To enable a quantitative interpretation of such time-series (e.g., Linde et al., 2011), it is necessary to fill the gaps and remove the interference signals.

The initial step of the filtering requires separating the clean data (training data) from data gaps or interference-affected data (data to be filtered). Although data gaps are easily identified, the identification of interferences is case-dependent. In our example, given the large electrical currents injected in the medium from different locations in a short time period, interferences are identified based on the absolute value of the temporal derivative of the time series. We define a threshold above which the self-potential signal is considered as affected by interferences. To clearly delineate these periods, four time steps before and after are also tagged as affected.

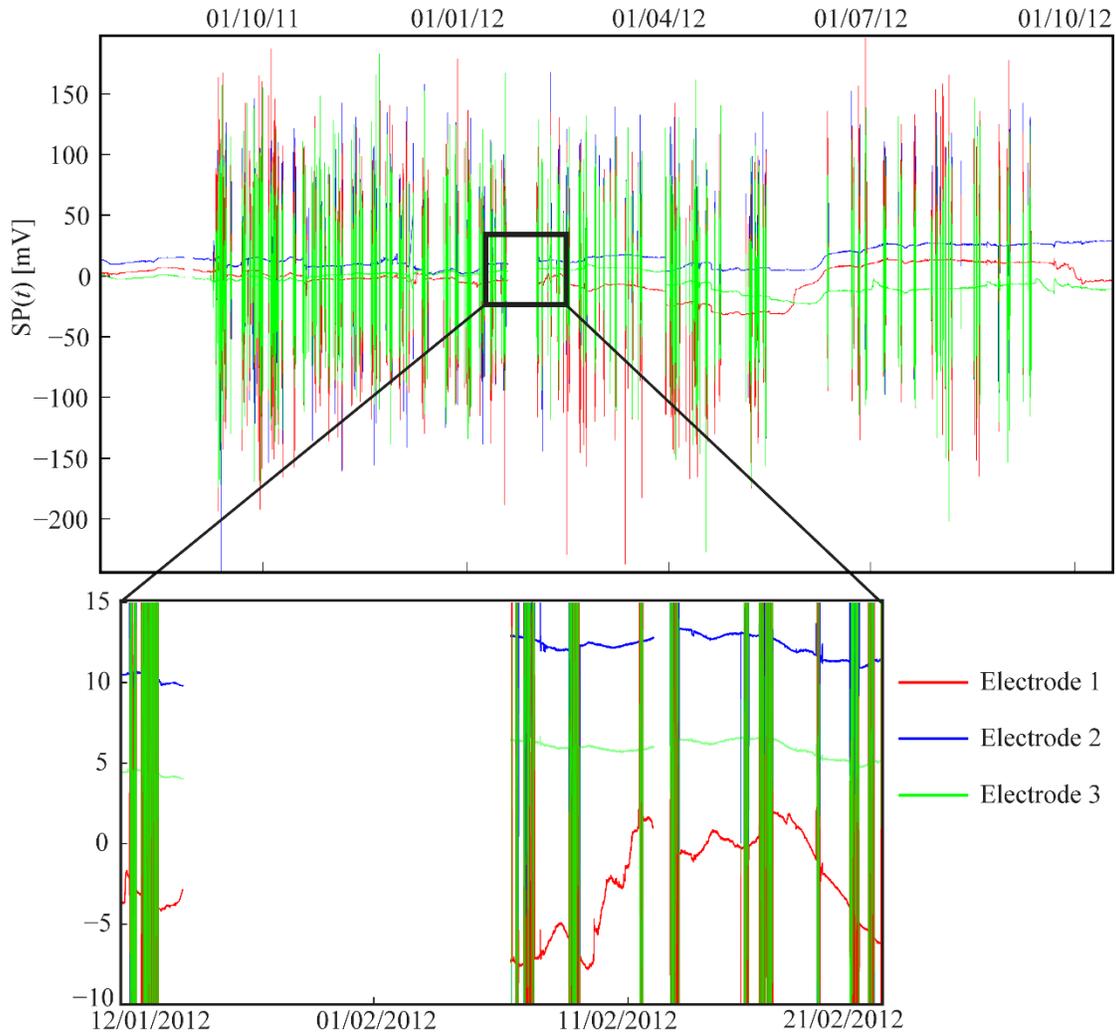

**Figure 2.** Self-potential data series that are strongly affected by data gaps and interferences from electrical resistivity tomography campaigns at the site.

### 3.2. Filtering results

Our filtering method was applied to process the test data set. We used a distance threshold of $\tau=0.02$ and $n=10$ neighbors for each of the 3 variables, meaning that we consider multivariate patterns of 30 time stamps taken simultaneously. Fig. 3 shows the results of 10 stochastic realizations. The zoomed area is the same as in Fig. 2. Convergence is assessed by analyzing the mean variation-based distance (eq. 1) for one realization. After 60 iterations, this distance stabilizes at less than 1% of the initial value, thereby indicating convergence. Similar convergence rates are observed for all realizations.

Visually, the filtered interference-affected periods appear to retain the trend characteristics of the underlying signal and are not artificially smoothed. Furthermore, the results capture the specific characteristics of each electrode signal, including features that could be considered as noise such as spikes of low amplitude. These spikes are present in many parts of the training data set. Since they have not been removed explicitly prior to filtering, they are an integral part of the signal to be reconstructed. For the reconstruction of the large gap located around the date 01/02/2012 (Fig. 3), the realizations provide different estimates. This variability allows quantifying the uncertainty related to the large gap where none of the three variables are informed. If data for at least one variable were available in this gap, it would guide the algorithm and result in lower variability in the results. The next sections quantitatively analyze the performance of our filtering method.

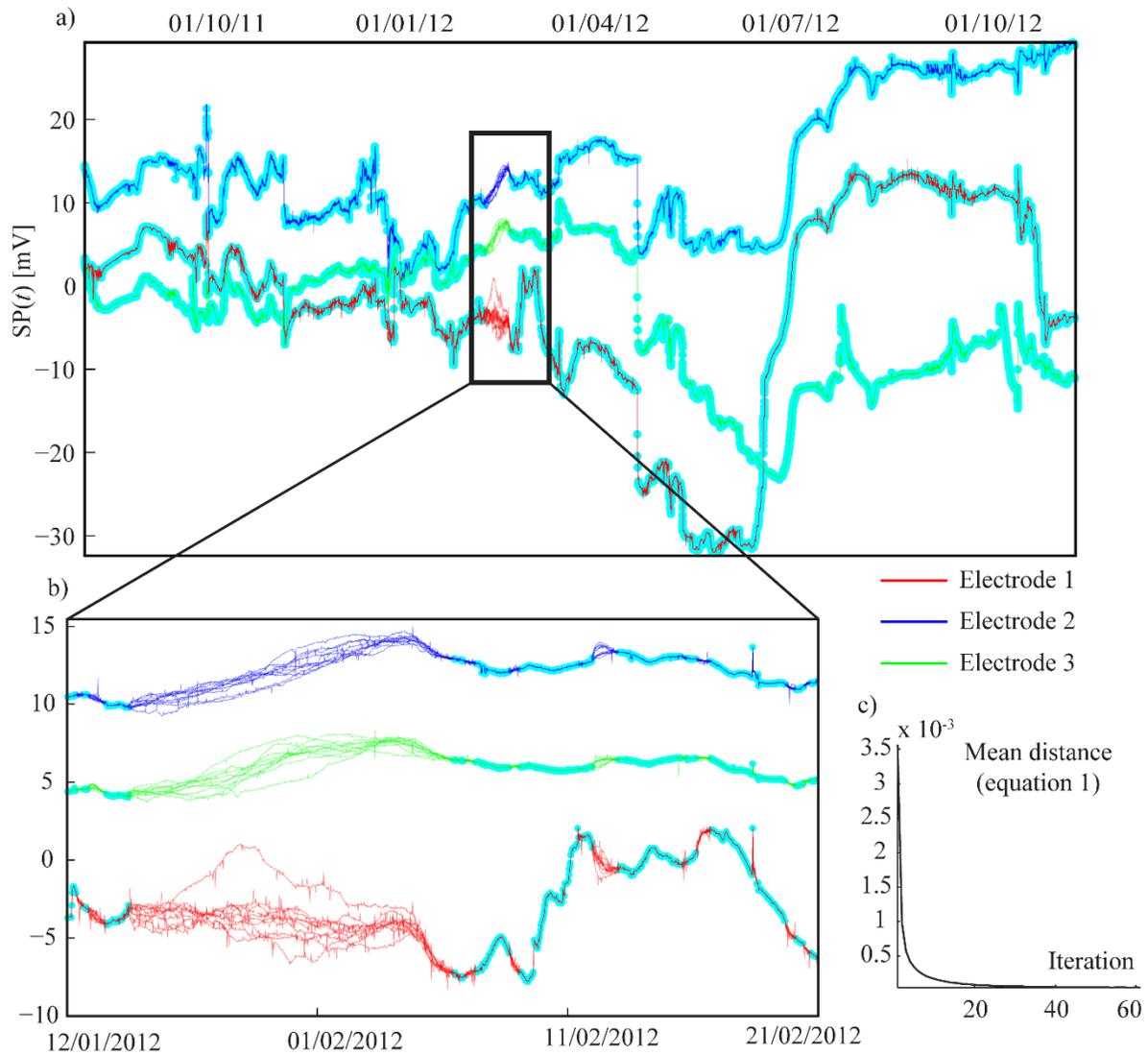

**Figure 3.** a) 10 realizations of filtering results. Data shaded in blue are interference-free, hence not filtered and used as training data. Data not shaded in blue are filtered results. b) Zoomed time period. c) Convergence plot, showing the mean variation-based distance (equation 1) in the same units as the filtered variable.

### 3.3. Conservation of scaling behavior

One important characteristic of our filtering method is that the results respect the variability of the original signal. To quantitatively compare this variability across a range of scales, we analyze the scaling behavior of the time series by determining the number of self-similar objects that are needed to cover an entire system as a function of the object size. This is a classical measure which is used to compute the fractal dimension of data when displayed on a log-log graph (Mandelbrot, 1967). Although here we do not compute fractal dimensions nor assume any fractal behavior in the dataset studied, we use the log-log plot as a tool to analyze the behavior of the filtered signal across a range of scales. The length of the training data set (only parts of the data set not affected by gaps/interferences) and filtered signals (only simulated parts) are computed for all three electrodes. The result is the total length of the signal ($L$, normalized by the maximum length found, $L_{max}$) as a function of the size of the ruler (Fig. 4a). It shows that the filtered outputs preserve the scaling characteristics of the data. As a comparison, Figure 4a shows the same analysis based on gap filling using spline interpolation, where the scaling properties are not reproduced because of the smoothing properties of splines.

The interdependence between the signals of the three electrodes is shown in Figure 4b where the scatter plots between the signals of electrodes 1 and 3 are compared for the training data and for the reconstructed values. While not identical, these relationships are similar, especially when considering that they are based on different parts of the signal: the training data set characterizes only the periods unaffected by interferences/gaps, while the other plots characterize only the affected periods. When using a spline interpolation, the relationships between electrodes are not accurately reproduced due to the smoothing nature of the interpolator. Similar observations are made with the other pairs of electrodes.

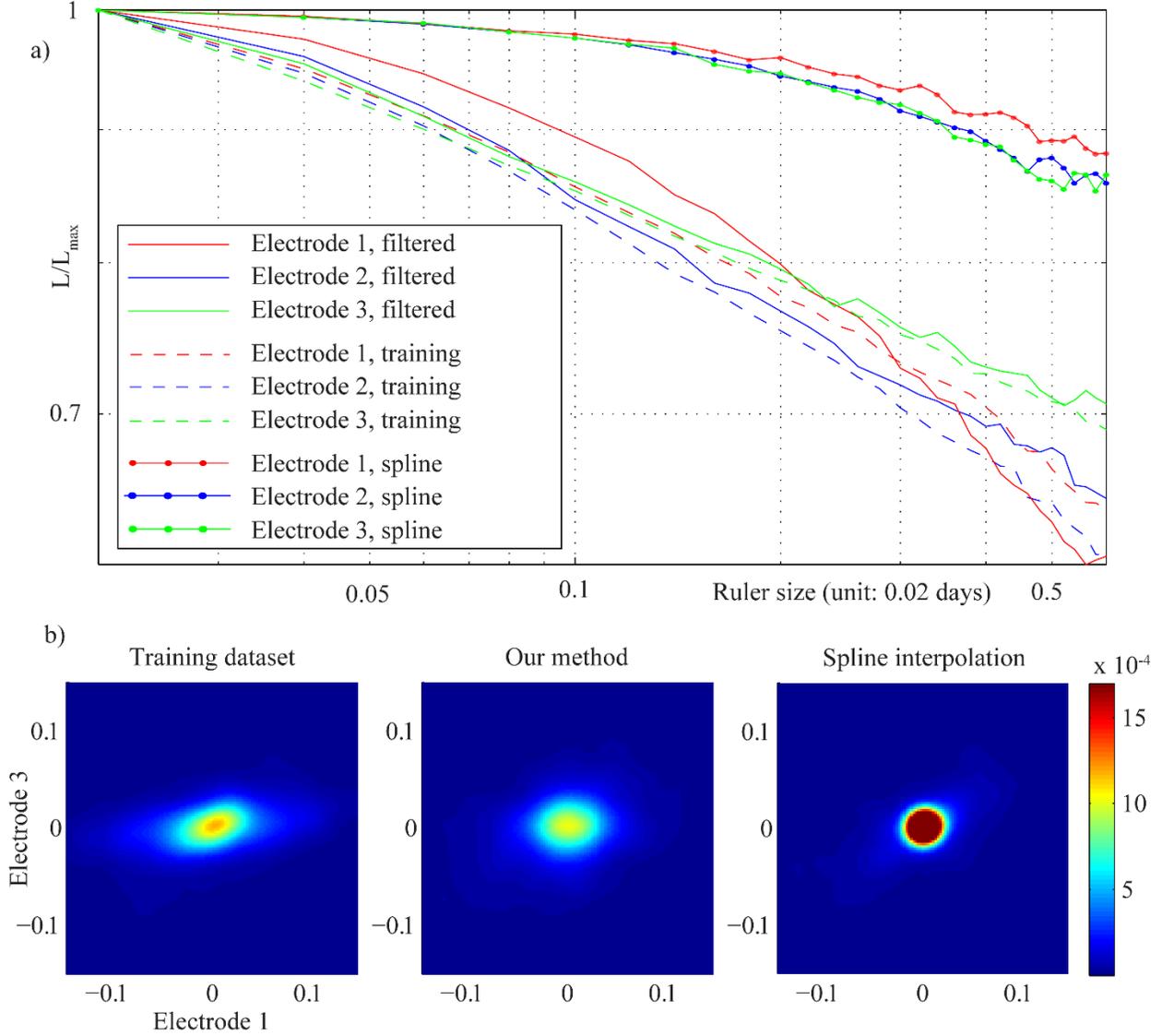

**Figure 4.** Statistical evaluation of the reconstructed signal. a) Scaling analysis of the signals. b) Dependences between electrodes 1 and 3 in the simulated signal and in the training data (density smoothing applied to scatter-plots). To avoid spurious correlations due to the non-stationarity of the signal, we consider data that are detrended using a moving average (window of 11 time stamps), and therefore centered on zero.

### 3.4. Cross-validation

In the field example considered, we don't know what data would have been measured in the absence of gaps and interferences. To quantitatively assess the performance of our method, we apply a cross-validation test that consists in adding interferences to a randomly chosen part of the training data with duration of 40 hours (the added interferences are independent for each electrode), and using the remaining part of the training data for validation. The approach is repeated 1000 times to derive statistics on the error reduction for different types of interferences. The first interference type is a uniformly distributed white noise of a magnitude 10 times larger than the standard deviation of the data. The second interference type consists of a highly asymmetric noise structure constructed by taking the absolute value of a normal distribution with a magnitude 10 times larger than that of the data. Note that adding asymmetric interferences include a bias, and are therefore even more challenging than missing data. The filtering results are compared to the original data. The results (see supplementary figure SF1 and table S2 for detailed values) show that our filtering method significantly reduces the magnitude of the errors in all cases, with standard deviations of errors being reduced from 32-90 mV before filtering to values of 3.5 mV after filtering. At the same time the procedure removes bias in the data: in the case of asymmetric interferences, the mean error is reduced from over 100 mV to under 1 mV.

## 4. Discussion and Conclusions

A novel time series processing method based on pattern matching has been presented which can remove interferences-affected artifacts and data gaps. While it has been tested on a case that considers several variables, it is equally applicable to single-variable time series. When applied to multivariate time-series, it preserves linear and non-linear dependences between the variables.

Several statistical measures show that it accurately reconstructs geophysical signals that are affected by measurement issues such as data gaps or interferences. In contrast to common filtering methods such as low-pass filters, our method does not systematically smooth the data. It preserves the intrinsic structure of each time series signal, including possible measurement noise present in the training data (parts of the signal defined as interference-free). The method performs well to correct the structure of the signal in the problematic areas, it does not remove measurement noise that is present in the training data, and therefore is not appropriate to separate an underlying signal from a noise component (e.g., Hristopulos et al., 2007).

The method is in theory applicable to affected periods or gaps of any size, however large gaps where none of the variables are informed (which we systematically used in our tests) are clearly the most challenging case, resulting in increased uncertainty. As with all training image-based techniques, the main limitation of the approach is the availability of training data, which should be large enough to be able to approximate ergodic behavior. In our test case the time series is long, comprising over 100,000 time stamps providing ample statistical basis for resampling a large variety of temporal patterns. However, care should be taken when applying our method to time series that are too short to represent the full variability of the signal (e.g. a limited number of time stamps for a very variable and non-stationary signal). Similarly, our method is not able to generate extreme values beyond the range of what is found in the training data since it resamples values from the training dataset (Oriani et al., 2014). Another limitation isthat the interference-affected periods have to be identified prior to applying the filtering. This can be done as in our test case by assuming a minimum level of smoothness in the signal over which it is considered that an interference is occurring. In other cases, these periods might have a known external cause (e.g. periodic magnetic interference) whose recurrence is identifiable.

# 5. Acknowledgements

The authors thank the Danish hydrological observatory HOBE for the access to the site and technical help. The computer code used in this study is available for academic purposes. It can be obtained from G. Mariethoz (gregoire.mariethoz@unil.ch) with a documentation and examples.

# 6. References

Bárdossy, A., Pegram, G., 2014. Infilling missing precipitation records – A comparison of a new copula-based method with other techniques. Journal of hydrology 519, Part A(0) 1162-1170.

Broersen, P.M.T., 2006. Automatic spectral analysis with missing data. Digital Signal Processing: A Review Journal 16(6) 754-766.

Hristopulos, D.T., Mertikas, S.P., Arhontakis, I., Brownjohn, J.M.W., 2007. Using GPS for monitoring tall-building response to wind loading: Filtering of abrupt changes and low-frequency noise, variography and spectral analysis of displacements. GPS Solutions 11(2) 85-95.

Jougnot, D., Linde, N., Haarder, E., Looms, M., in press. Monitoring of saline tracer movement with vertically distributed self-potential measurements at the HOBE agricultural site, Voulund, Denmark. Journal of Hydrology, 521, 314-327.

Kim, T.W., Ahn, H., 2009. Spatial rainfall model using a pattern classifier for estimating missing daily rainfall data. Stochastic Environmental Research and Risk Assessment 23(3) 367-376.

Kondrashov, D., Ghil, M., 2006. Spatio-temporal filling of missing points in geophysical data sets. Nonlinear Processes in Geophysics 13(2) 151-159.

Linde, N., Doetsch, J., Jougnot, D., Genoni, O., Dürst, Y., Minsley, B.J., Vogt, T., Pasquale, N., Luster, J., 2011. Self-potential investigations of a gravel bar in a restored river corridor. Hydrology and Earth System Sciences 15(3) 729-742.

Mandelbrot, B., 1967. How Long Is the Coast of Britain? Statistical Self-Similarity and Fractional Dimension. Science 156(3775) 636-638.

Mariethoz, G., Renard, P., Straubhaar, J., 2010. The direct sampling method to perform multiple-point geostatistical simulations. Water Resources Research 46(11).

Oriani, F., Straubhaar, J., Renard, P., Mariethoz, G., 2014. Simulation of rainfall time series from different climatic regions using the direct sampling technique. Hydrology and Earth System Sciences 18(8) 3015-3031.

Paparella, F., 2005. Filling gaps in chaotic time series. Physics Letters, Section A: General, Atomic and Solid State Physics 346(1-3) 47-53.

Revil, A., Jardani, A., 2013. The self-potential method: Theory and applications in environmental geosciences. Cambridge University Press.

Ruiz-Alzola, J., Alberola-López, C., Westin, C.F., 2005. Kriging filters for multidimensional signal processing. Signal Processing 85(2) 413-439.

Schoellhamer, D.H., 2001. Singular spectrum analysis for time series with missing data. Geophysical Research Letters 28(16) 3187-3190.

Shannon, C., 1948. A mathematical theory of communication. The Bell system technical journal(27) 379-423.

Smith-Boughner, L.T., Constable, C.G., 2012. Spectral estimation for geophysical time-series with inconvenient gaps. Geophysical Journal International 190(3) 1404-1422.

Teegavarapu, R.S.V., Chandramouli, V., 2005. Improved weighting methods, deterministic and stochastic data-driven models for estimation of missing precipitation records. Journal of hydrology 312(1-4) 191-206.

Wang, G., Garcia, D., Lui, Y., de Jeu, R., Dolman, A., 2012. A three-dimensional gap filling method for large geophysical datasets: Application to global satellite soil moisture observations. Environmental Modelling & Software 30 139-142.

Žukovič, M., Hristopulos, D.T., 2013. Reconstruction of missing data in remote sensing images using conditional stochastic optimization with global geometric constraints. Stochastic Environmental Research and Risk Assessment 27(4) 785-806.